\documentclass[12pt,eqsecnum,preprint]{aastex}

\begin{document}

\title{3C 216: A Powerful FRII Seyfert 1 Galaxy}
\author{Brian Punsly} \affil{4014 Emerald Street No.116,
Torrance CA, USA 90503 and International Center for Relativistic
Astrophysics, I.C.R.A.,University of Rome La Sapienza, I-00185
Roma, Italy} \email{brian.m.punsly@L-3com.com or
brian.punsly@gte.net}

\begin{abstract}3C 216 has a weak accretion flow
luminosity, well below the Seyfert1/QSO dividing line, weak broad
emission lines (BELs) and powerful radio lobes. As a consequence
of the extreme properties of 3C 216, it is the most convincing
example known of an FR II radio source that is kinetically
dominated: the jet kinetic luminosity, $Q$, is larger than the
total thermal luminosity (IR to X-ray) of the accretion flow,
$L_{bol}$. Using three independent estimators for the central
black hole mass, we find that the jet in 3C 216 is very
super-Eddington, $3.3 L_{Edd}<\overline{Q}< 10 L_{Edd}$, where
$\overline{Q}$ is the long term time averaged $Q(t)$, calculated
at 151 MHz. It is argued that 3C 216 satisfies the contemporaneous
kinetically dominated condition, $R(t)\equiv Q(t)/L_{bol}(t)>1$,
either presently or in the past based on the rarity of
$L_{bol}>L_{Edd}$ quasars. The existence of $R(t)>1$ AGN is a
strong constraint on the theory of the central engine of FRII
radio sources.
\end{abstract}


\section{Introduction}This article describes the extreme
properties of the radio source 3C 216 and the implications to the
theory of the central engines of FR II radio sources. It has
powerful lobe emission that implies an enormous time averaged jet
kinetic luminosity ($\overline{Q}\approx 1.5\times
10^{46}\mathrm{ergs/s}$), narrow emission lines ($\sim 1800$
km/s), and it has a very low accretion flow thermal luminosity,
$L_{bol}< 10^{45}\mathrm{ergs/s}$ (below the Seyfert 1/QSO
dividing line). If it were not for the powerful jet seen in a
pole-on orientation, 3C 216 would be a very ordinary Seyfert 1
galaxy: $L_{bol}\gtrsim 10^{44} \mathrm{ergs/s}$ and
$L_{bol}/L_{Edd}\approx 0.05$. This object is unusual in that
Seyfert 1 galaxies are typically radio quiet and $\overline{Q}$ in
3C 216 is an order of magnitude larger than that of any other jet
known to emanate from a Seyfert 1 nucleus \citep{rod05}. The
Mid-IR sample of \citet{ogl06} made it clear that many FR II
narrow line radio galaxies (NLRGs) had no hidden quasar, but it
was unclear if there was a faint central broad line AGN and how
many (if any) of the objects had unobstructed lines of sight to
the accretion disk. The blazar core in 3C 216 makes the
orientation unambiguous, there is a direct line of sight to the
BEL region (BLR, hereafter) and accretion disk. The Ly$\alpha$
line is broad with FWHM = 2600 km/sec, in an HST spectrum,
\citet{wil95}, and anything over 1500 km/sec is "out of family"
with even the most luminous, high redshift NLRGs (Patrick McCarthy
private communication 2006). Thus, 3C 216 is a BEL AGN and is
clearly distinct from a typical obscured FRII NLRG \citep{ant93}.
These properties allow one to assert that the standard
astronomical estimators indicate a very small central black hole
mass $M_{bh}\sim 10^{7}M_{\odot}$ and the jet has a time averaged
Eddington luminosity, $3.3
<\overline{Q}_{Edd}\equiv\overline{Q}/L_{Edd}< 10$.
\par It is not known how
powerful an FRII jet can be relative to the thermal radiative
luminosity from accretion, $L_{bol}$. Understanding the limits of
jet power can help reveal the physical nature of the quasar
central engine. The primary conundrum posed by such a task is that
$L_{bol}$ and $Q$ need to be estimated contemporaneously and there
is no reliable method for measuring the instantaneous $Q(t)$ (see
section 2). The only reliable estimates that we have are for
$\overline{Q}$. Some $\overline{R}\equiv \overline{Q}/L_{bol}>1$
sources were found in \citet{pun06}. Unfortunately the
$\overline{Q}$ estimate is not contemporaneous with the $L_{bol}$
data, so one can not say if these sources presently satisfy or
ever satisfied $R(t)\equiv Q(t)/L_{bol}(t)>1$, (the energy flux
emerging from the AGN in the jet exceeds that of the accretion
disk radiation at the time, t). The fact that 3C 216 is so
extreme, $\overline{Q}> 3.3 L_{Edd}$, allows us to overcome the
limitations of using $\overline{Q}$ and deduce that episodes of
$R(t)>1$ exist in FR II radio sources (in this paper, we adopt the
following cosmological parameters: $H_{0}$=70 km/s/Mpc,
$\Omega_{\Lambda}=0.7$ and $\Omega_{m}=0.3$ in the calculations).
There is no evidence that any quasar accretion flow can create an
equal level of $L_{bol}$ for the lifetime of a radio source, or
even a short period of time for that matter. Thus, $R(t)>1$,
either presently or in the past based on the rarity of
$L_{bol}>L_{Edd}$ quasars. This makes the most convincing argument
to date for a kinetically dominated jet, $R(t)>1$, in an FR II
radio source.
\section{Estimating The Time Averaged Kinetic Luminosity} Even though one can use the
jet emission from the parsec scale radio core to estimate, $Q$
more contemporaneously with the accretion flow emission as in
\citet{cel97}, such estimates are prone to be very inaccurate. A
review of these issues can be found in \citet{tin05} and a
published example in which the $Q$ is apparently miscalculated
using radio core properties by three orders of magnitude is
discussed explicitly. Thus, the most accurate estimates of $Q$
should rely on isotropic properties. An isotropic method that
allows one to convert 151 MHz flux densities, $F_{151}$ (measured
in Jy), into estimates of $\overline{Q}$ (measured in ergs/s), was
developed in \citet{wil99,blu00}, the result is captured by the
formula derived in \citet{pun05}:
\begin{eqnarray}
&& \overline{Q} \approx 1.1\times
10^{45}\left[(1+z)^{1+\alpha}Z^{2}F_{151}\right]^{\frac{6}{7}}\mathrm{ergs/s}=
1.51\times 10^{46}\mathrm{ergs/sec}\;,\\
&& Z \equiv 3.31-(3.65) \nonumber \\
&&\times\left(\left[(1+z)^{4}-0.203(1+z)^{3}+0.749(1+z)^{2}
+0.444(1+z)+0.205\right]^{-0.125}\right)\;,
\end{eqnarray}
where $F_{151}$ is the total optically thin flux density from the
lobes (i.e., \textbf{no contribution from Doppler boosted jet or
the radio core}). We define the radio spectral index, $\alpha$, as
$F_{\nu}\propto\nu^{-\alpha}$. Alternatively, one can also use the
independently derived isotropic estimator from \citet{pun05}
\begin{eqnarray}
&&\overline{Q}\approx
5.7\times10^{44}(1+z)^{1+\alpha}Z^{2}F_{151}\,\mathrm{ergs/s}=1.41\times
10^{46}\mathrm{ergs/s}\;,\quad\alpha= 1.09\;.
\end{eqnarray}
\par In order to utilize the LHS of these equations, one must extricate the
diffuse lobe emission from the Doppler boosted core of 3C 216.
Fortunately, there are many high dynamic range radio maps of 3C
216 \citep{fej92,bar88,pea85}. The 408 MHz MERLIN map of
\citet{fej92} shows a very luminous halo enveloping the core. The
blazar-like, variable, flat spectrum radio core surrounded by a
diffuse halo is the classic configuration of an FR II AGN viewed
close to the jet axis in the "standard model" \citep{ant93}. Using
the numerous photometric data points in NED and the core fluxes
from \citet{fej92,bar88,pea85}, one can fit the radio spectrum
from 86 MHz to 37 GHz very accurately with a simple two component
model. The core is variable and flat spectrum with an average flux
density of about 0.9 Jy up to 37 GHz, where the spectrum starts
to turnover. The other component is very steep with $\alpha\approx
1.09$. This is the halo emission which achieves a flux density of
26.4 mJy at 151 GHz. The estimators (2.1) and (2.3) are most
accurate for steep lobe emission as in 3C 216 \citep{wil99,pun05}.
\section{Estimating the Bolometric Luminosity} The total
bolometric luminosity of the accretion flow, $L_{bol}$, is the
thermal emission from the accretion flow, including any radiation
in broad emission lines from photo-ionized gas, from IR to X-ray.
To estimate, $L_{bol}$, we use the composite SED in table 2 of
\citet{pun06}, which see for details. This spectrum, in
combination with the BELs, represents the ``typical'' radiative
signature of a strong accretion flow onto a black hole. This
signature is empirical and it is independent of all theoretical
models of the accretion flow. If $L(\nu)_{\mathrm{obs}}$ is the
observed spectral luminosity at the AGN rest frame frequency,
$\nu$, then $L_{bol}$ is estimated as
\begin{eqnarray}
&& L_{bol}=1.35\frac{\nu L(\nu)_{\mathrm{obs}}}{\nu
L(\nu)_{\mathrm{com}}}\times 10^{46}\mathrm{ergs/sec}< 1.01 \times
10^{45} \mathrm{erg/s}\;,
\end{eqnarray}
 where $L(\nu)_{\mathrm{com}}$ is the spectral luminosity from the composite SED.
\par The problem with applying the LHS of (3.1) to 3C 216 is that one must
subtract off the high frequency tail of the synchrotron jet. The
optical component is prodigious compared to the accretion disk.
The optical emission in 3C 216 is marked by extreme variability,
21\% optical polarization, \citet{imp91}, and a steep power law in
the optical, all confirming that the jet emission dominates. The
best way to dig through this powerful jet emission down to the
hidden accretion disk luminosity is to observe the object many
times and hopefully measure 3C 216 in a low state. A low state was
observed in \citet{gel94}, a flux density of $1.7\times 10^{-17}
\mathrm{ergs/s/\AA}$ at $4190 \AA$ in the QSO frame. This result
inserted into the LHS of (3.1), yields the numerical estimate on
the RHS of (3.1). However, the optical spectral index is very
steep, $\alpha=2.65$, so it is likely that the the underlying
continuum is still hidden by the jet and this is verified by the
BEL strengths discussed below.
\par Another method that is commonly used
for estimating $L_{bol}$ in blazars is to compare the line
strengths to a composite SED \citep{cel97,wan04}. From
\citet{law96}, the line strengths for MgII and H$\beta$ are
$8.7\times 10^{41}\mathrm{erg/s}$ and $7.6\times
10^{41}\mathrm{erg/s}$, respectively. Since the line strengths are
so weak, subtraction of the narrow line component from H$\beta$ to
get the broad component, H$\beta_{BC}$, is essential for accuracy.
We employ the prescription commonly used, subtract 0.1 of the
[OIII] 5007 narrow line (which is the typical ratio of line
strengths in NLRGs) from the H$\beta$ \citep{mcc93,mar06}. The
H$\beta_{BC}$ line strength is $4.6\times 10^{41}\mathrm{erg/s}$
and the F(H$\beta_{BC}$) = 1685 km/s (where F() means "FWHM of" ,
hereafter). Using the estimators in eqn(1) of \citet{wan04},
$L_{bol}\approx 1.2 \times 10^{44} \mathrm{erg/s}$. There is no
similar standard prescription for the narrow line subtraction in
MgII \citep{mcc04}. The MgII line, without the narrow line
subtraction, can still provide a useful upper bound, $L_{bol} <
1.4 \times 10^{44} \mathrm{erg/s}$ \citep{wan04}. Note that the
BEL estimates of $L_{bol}$ and (2.1)- (2.3), imply that
$\overline{R}\equiv \overline{Q}/L_{bol}> 100$!
\section{Estimating the Black Hole Mass}
 We estimate $M_{bh}$ by two independent methods:, first we
 use the continuum flux density and the FWHM of the low ionization
BELs and secondly from the host galaxy bulge luminosity - mass
relation. The estimate, (4.1), is derived from the virial
assumption in \citet{ves06} applied
 to a sample of 32 reverberation measurements of AGN that provide the velocity ($\sim$ FWHM)
and the radius of the BLR (time-delay times c). A best fit BLR
radius - continuum luminosity curve is created for these AGN.
Using this relation, $L_{\lambda}(5100 \AA)$ of a single epoch
observation can be used as a surrogate for the radius of the BLR.
Inserting the continuum flux from \citet{gel94} into eqn(5) of
\citet{ves06} yields,
\begin{eqnarray}
 && M_{bh}(H \beta)= 10^{6.91 \pm
0.02}\left(\frac{F(H\beta_{BC})}{1000
\mathrm{km/s}}\right)^{2}\left(\frac{\lambda L_{\lambda}(5100
\AA)}{10^{44}\mathrm{ergs/s}}\right)^{0.50} < 2.6 \times
10^{7}M_{\odot}\;.
\end{eqnarray}
The assumption behind this method of taking the FWHM and
$L_{\lambda}(5100 \AA)$ from different observations is the
following. $L_{\lambda}(5100 \AA)$ in (4.1) represents the
ionizing continuum flux. This flux originates in the accretion
disk, since the optical flux from the jet is beamed orthogonal to
and away from the low ionization BEL gas \citep{bro86}. The accretion disk
$L_{\lambda}(5100 \AA)$ is assumed to be fairly constant on the
time scale of years and it is the masking radiation of the
synchrotron jet that is changing in intensity. The low states
gives the most accurate upper bound on $L_{\lambda}(5100 \AA)$.
The inequality sign in (4.1) is there because we don't know if the
observed spectrum has some contribution from the jet that
increases $L_{\lambda}(5100 \AA)$ above the actual ionizing source
value. This is a very low mass for an FRII radio source so it is
fortunate that we have a second mass estimate for verification.
Using F(Mg II) = 1790 km/s from \citet{law96} and the continuum
flux density at $3000\AA$ (extrapolated from the $4190\AA$ value
in \citet{gel94}, with $\alpha=0.7$ as in the SED of
\citet{pun06}) in the estimator of \citet{kon06} provides a second
estimate:
\begin{eqnarray}
&& M_{bh}(\mathrm{MgII})= 3.4\times 10^{6}
\left(\frac{F(\mathrm{MgII})}{1000
\mathrm{km/s}}\right)^{2}\left(\frac{\lambda L_{\lambda}(3000
\AA)}{10^{44}\mathrm{ergs/s}}\right)^{0.58 \pm 0.10}< 1.7 \times
10^{7}M_{\odot}\;.
\end{eqnarray}
The results in (4.1) and (4.2) are consistent indicating that the
virial estimates are tracking the gravitational potential.
\par The most obvious source of error in the virial mass estimates is the effect of
near pole-on orientation. We quantify the orientation correction
by defining a sample of extremely superluminal quasars with a
projected apparent velocity on the plane of the sky of
$\beta_{a}\geq 10$ (viewed virtually pole-on) and comparing their
FWHM to "average" off-axis lines of sight, steep spectrum, radio
loud quasars (SSQSRs). The $\beta_{a}\geq 10$ quasars have mean
$\overline{F}$(H$\beta$) and $\overline{F}$(Mg II) of $3955\pm
1364$ km/s, and $4125\pm 1997$ km/s, respectively and the SSQSRs
have mean FWHM of $5772 \pm 3533$ and $6398\pm 2668$ km/s
\citep{pun07}. If 3C 216 is typical of the $\beta_{a}\geq 10$
extreme blazar population then the factor which $M_{bh}$ has been
underestimated in (4.1) and (4.2) due to orientation effects can
be approximated by $0.47=(3955/5772)^{2}$ for H$\beta$ and
$0.42=(4125/6398)^{2}$ for MgII. This analysis of $\beta_{a}\geq
10$ quasars is in line with the orientation correction factor for
flat spectrum quasars (FSQSRs) of 0.5 determined in \citet{osh02}
and the inference from the sample of \citet{jar06} is a correction, 
$(\overline{F}(FSQSRs)/\overline{F}(SSQSRs))^{2}=(4990/6464)^{2}=0.6$.
\par In the last section, it was noted that the line strengths
and the continuum spectral index indicate that $L_{bol}$ is
overestimated in (3.1) by a factor $\sim 8$. Thus the effects of
overestimating $L_{bol}$ and the pole-on orientation almost cancel
out in (4.1) and (4.2). The estimates corrected for orientation
and and continuum luminosity are:
\begin{eqnarray}
&& M_{bh}=(\frac{1}{0.47\sqrt{8}})2.6 \times 10^{7}M_{\odot}=
2.0\times 10^{7}M_{\odot}\;,\; H\beta\\
&& M_{bh}=(\frac{1}{0.42(8)^{0.58}})1.7 \times 10^{7}M_{\odot}=
1.2\times 10^{7}M_{\odot}\;,\; \mathrm{Mg II}\;,
\end{eqnarray}
\par The estimates in (4.3) and (4.4) are consistent with each other, but
 it is best to verify the result with a completely independent method such as
 one based on the galactic bulge luminosity. The HST image
has been given in \citet{dev97,leh99}. This source is not trivial.
The high resolution contour map in \citet{dev97} shows a strong
optical jet directed to the southeast that is coincident with the
VLBI jet, \citet{bar88,fej92}, representing $>55\%$ of the
extended optical flux. There is a region of diffuse extended
luminosity that is coincident with the northern radio lobe in the
1.6 GHz MERLIN map of \citet{fej92}. To the east, there is excess
emission enveloping the jet that is likely to be narrow line
emission \citep{mcc93}. The host galaxy bulge represents only
$12\%$ of the total extended optical flux which equates to a k
corrected value of $M_{R}\approx -21.3$ \citep{fuk95}. Inserting
$M_{R}=-21.3$ into the best fit estimator derived from 72 AGN in
table 3 of \citet{dun02}, yields $M_{bh}= 3.5 \times
10^{7}M_{\odot}$ in close agreement with (4.3) and (4.4).

\section{A Super Eddington Jet} The $M_{bh}$ estimates combined with
the $\overline{Q}$ estimates in section 2, imply,
$3.3<\overline{Q}_{Edd}< 10$. A value of $\overline{Q}_{Edd}>3.3$
suggests that even if the jet central engine is currently in a low
state then at some epoch during the lifetime of the source it must
have had $R(t)>1$. For example, if 3C 216 is not now nor has ever
been kinetically dominated then the on average
$\overline{L}_{bol}> 3.3 L_{Edd}$ for the lifetime of the source.
However, since $\overline{Q}=(1/T)\int_{0}^{T}Q(t) dt$, the peak
values of the instantaneous $Q(t)$,
$\mathrm{max}_{t}\left[Q(t)\right]> 3.3 L_{Edd}$. Thus, one
expects that $\mathrm{max}_{t}\left[ L_{bol}(t)\right]$ would have
to exceed $3.3 L_{Edd}$ by a significant amount at certain epochs
in order for the 3C 216 to have always been in an $R(t)<1$ state.
This is inconsistent with the magnitudes of the peaks in the duty
cycle of quasar $L_{bol}(t)/L_{Edd}$ based on our current
knowledge, as discussed below.
\par Applying (4.1) to a
sample of 232 broad line, F(H$\beta) > 2000\mathrm{km/s}$, AGN
(138 radio loud and 94 radio quiet) that have been used in
published estimates of $M_{bh}$ in
\citet{ves06,guu01,bas04,mcc01,osh02,wan04,tin05} reveal no
$L_{bol}> L_{Edd}$ sources. In regards to narrow line AGN, it was
noted in \citet{mcc02} that many have broad UV lines which
indicates that H$\beta$ is not always an accurate tracker of the
gravitational potential. This claim is borne out by the 13 narrow
line objects with $L_{bol}> L_{Edd}$ in the combined, broad and
narrow line, sample of 268 AGN from the references above. Archival
UV emission lines are available for 9 of the 13 objects. All 9
show broad UV lines. When F(MgII) is inserted into to (4.2) and
F(CIV) is inserted into eqn(7) of \citet{ves06}, these sources are
transformed into sub-Eddington AGN. The best available data
indicates that $L_{bol}> L_{Edd}$ is a rare or nonexistent phase
in the duty cycle of a quasar activity. The most likely
explanation of the $Q_{Edd}\sim 5$ estimates is that 3C 216
experienced epochs in which $R(t)\sim 5-10$ as opposed to the
alternative explanation that it had protracted phases in which
$L_{bol}(t)/L_{Edd}\sim 5-10$.

\section{Discussion} The discussion of 3C 216 demonstrates that FR II
sources seem to exist in a state of $R(t)>1$. The
$\overline{Q}_{Edd}\sim 5$ jet in 3C 216 must satisfy
$\mathrm{max}_{t}\left[R(t)\right]>1$, or else the accretion disk
would episodically be in an un-physical state of
$L_{bol}(t)/L_{Edd}\gg 1$. This results is important because it
reflects on the physical nature of the central engine.
\par As a consequence of the near independence of the UV spectrum on the radio state, \citet{dev06,cor94},
it has been argued that the black hole and not the accretion disk
is the power source for FR II jets \citep{sem04}. However, a large
scale magnetic flux trapped in the vortex inside the accretion
disk can tap the black hole rotational energy as seen in the MHD
numerical simulations of \citet{haw06,mck05,gam04,sem04}. Yet, it
is not a trivial matter to obtain the $R>1$ condition. The
simulations of \citet{haw06,mck05,gam04} are very similar and rely
on the magnetic flux threading the event horizon of the black hole
to source the jet. These simulations are hard pressed to achieve
$R\gtrsim 0.1$ for accretion disk parameters representative of
quasars (i.e., an accretion disk radiating near its maximum thin
disk efficiency, $\epsilon > 0.2$, \citet{elv02}). This derives
from the fact that the simulations of \citet{haw06,mck05,gam04} do
not produce a net extraction of black hole energy, the black hole
actually absorbs energy during the simulation. By far the largest
R values occur in the new simulations of \citet{haw06} that are
parameterized by the black hole spin parameter, $a/M=0.99$. For
accretion disks with quasar-like parameters (see table 4 of
\citet{haw06}), $R=0.8$. However, this is an overestimate of R
that is driven by the numerical resolution of the model (McKinney
private communication 2006). Increasing the resolution in the
numerical mesh near the equatorial plane allows short wavelength
MRI modes to propagate in the accretion disk, which in turn
increases the accretion rate, so $R<0.5$ \citep{gam04,mck05}. Even
so, obtaining $R\gtrsim 0.1$, in \citet{haw06,mck05,gam04}
requires a very massive enveloping wind to collimate the jet. In
the $a/M=0.99$, $R=0.8$ model of \citet{haw06}, the wind
transports over $1 M_{\odot}/\mathrm{yr}$ outward at 0.3 c - 0.4 c
for FRII quasar parameters. Although it is never stated, the
thermal energy content must be enormous, $\sim 1\mathrm{GeV}$ in
order to balance the energy equation. If such massive winds
existed around FRII jets they would have certainly been
prominently detected in some frequency band even if they
radiatively cool to a more reasonable temperature. Furthermore, in
this numerical model the collimating wind transports far more
energy flux than $L_{bol}$ and $Q$ combined, a high price for
collimation.
\par Alternatively, the black hole energy extraction model of
\citet{pun01} and references therein that was numerically realized
in \citet{sem04} is based on large scale magnetic flux that
threads the equatorial plane of the ergosphere, the active region
outside of the black hole. The solutions attain a maximum $R$
value, $R_{\mathrm{max}}$, that was calculated for rapidly
rotating black holes, $a/M\approx 0.99$ in eqn (10.14) of
\citet{pun01}. Extending this result to lower spins
\begin{eqnarray}
R_{\mathrm{max}}\approx 11\frac{0.3}{\epsilon}(a/M)^{10.5}\; ,\,
0.9 <a/M<0.99\;,
\end{eqnarray}
where $\epsilon$ is the radiative efficiency of the accretion
flow, $L_{bol}\equiv \epsilon (dM/dt)$ and $(dM/dt)$ is the
accretion rate from the disk to the black hole. The maximum
efficiency from a thin disk is $\epsilon\approx 0.3$
\citep{nov73}. The exponent in (6.1) was estimated by
approximating the numerical data displayed in figure 10.13 of
\citet{pun01} in the range, $0.9 <a/M<0.99$. The coupling of $Q$
to $\epsilon$ occurs in (6.1) because the ram pressure of the
accretion flow regulates the maximum allowable magnetic pressure
and flux in the ergosphere. For objects below the Seyfert 1/QSO
dividing line, such as 3C 216, one might expect $a/M\approx 0.95$
and from \citet{nov73}, for a thin disk, $\epsilon=0.19$:
$R_{\mathrm{max}}\approx 10$ which is probably required
episodically in order to attain $\overline{Q}_{Edd}\sim 5$.

\begin{acknowledgements}
I am indebted to Ski Antonucci for his valuable comments that help
propel the development of this research. I am also thankful to Wim
Devries and Chris O'Dea for sharing their HST results.
\end{acknowledgements}

\end{document}